\documentstyle[a4,12pt]{article}
               
\newcommand{\nit}{\noindent}
\newcommand{\nl}{\newline}
\newcommand{\np}{\newpage}
\newcommand{\dsp}{\displaystyle} 
\newcommand{\vs}[1]{\vspace{#1 ex}}
\newcommand{\bfr}{\begin{flushright}}
\newcommand{\efr}{\end{flushright}}
\newcommand{\bc}{\begin{center}}
\newcommand{\ec}{\end{center}} 
\newcommand{\ben}{\begin{enumerate}}
\newcommand{\een}{\end{enumerate}} 

\newcommand{\be}{\begin{equation}}
\newcommand{\ee}{\end{equation}}
\newcommand{\ba}{\begin{array}}
\newcommand{\ea}{\end{array}}
\newcommand{\ct}{\cite}
\newcommand{\bit}{\bibitem}

\newcommand{\gam}{\gamma}
\newcommand{\del}{\delta}
\newcommand{\eps}{\epsilon}
\newcommand{\ve}{\varepsilon}
\newcommand{\zg}{\zeta}

\newcommand{\kg}{\kappa}
\newcommand{\lb}{\lambda}
\newcommand{\sg}{\sigma}

\newcommand{\og}{\omega}
\newcommand{\Gam}{\Gamma}
\newcommand{\Del}{\Delta} 
 
\newcommand{\Sg}{\Sigma}

\newcommand{\bfxi}{\mbox{{\boldmath $\xi$}}}

\newcommand{\lh}{\left(}
\newcommand{\rh}{\right)}

\newcommand{\pl}{\partial}

\newcommand{\bzg}{\overline{\zg}}

\begin{document}
               
\pagestyle{empty}
\begin{flushright} 
NIKHEF 99-018
\end{flushright}

\begin{center}
\Large{\bf{Gravitational waves and massless particle fields}} \\
\vs{3}
\large{J.W.\ van Holten$^*$}
\\ 
\vs{2}

\large{NIKHEF, Amsterdam NL} \\ 
\vs{2} 
{\tt t32@nikhef.nl} 
\\
\vs{10}

\small{\bf{Abstract}} \\
\end{center}
               
\nit
\footnotesize{
These notes address the planar gravitational wave solutions of 
general relativity in empty space-time, and analyze the motion of 
test particles in the gravitational wave field. Next we consider
related solutions of the Einstein equations for the gravitational 
field accompanied by long-range wave fields of scalar, spinor and 
vector type, corresponding to massless particles of spin $s = 
(0,\frac{1}{2},1)$. The motion of test masses in the combined 
gravitational and scalar, spinor or vector wave fields is 
discussed.   
}
\vfill 
\nit
\footnoterule 
\nit 
$^*$ Work performed as part of the research program of the 
Foundation for Fundamental Research of Matter (FOM).
               
\np
               
\pagestyle{plain}
               
\section{Planar gravitational waves}
                
{\bf a.\ Planar wave solutions of the Einstein equations} 
\vs{1} 

\nit  
Planar gravitational wave solutions of the Einstein equations have  
been known since a long time \ct{bondi}-\ct{ksmch}. In the following  
I discuss unidirectional solutions of this type, propagating along a  
fixed light-cone direction; thus the fields depend only on one of the  
light-cone co-ordinates $(u,v)$, here taken transverse to the  
$x$-$y$-plane:  
               \be  
               u = ct - z, \hspace{3em} v = ct + z.  
               \label{2}  
               \ee   
Such gravitational waves can be described by space-time metrics  
   \be  
   g_{\mu\nu} dx^{\mu} dx^{\nu} = -\, du dv - K(u,x,y) du^2 +  
   dx^2 + dy^2 = -\, c^2 d\tau^2,  
   \label{1}  
   \ee  
or similar solutions with the roles of $v$ and $u$ interchanged. If  
the space-time is asymptotically minkowskian. With the metric  
(\ref{1}), the connection co-efficients become  
\be  
\Gam_{uu}^{\;\;\;\;v} = K_{,u}, \hspace{2em}  
\Gam_{uu}^{\;\;\;\;x} = \frac{1}{2}\, \Gam_{xu}^{\;\;\;\;v} =  
\frac{1}{2}\, K_{,x}, \hspace{2em}  
\Gam_{uu}^{\;\;\;\;y} = \frac{1}{2}\, \Gam_{yu}^{\;\;\;\;v} =  
\frac{1}{2}\, K_{,y}.  
\label{3}  
\ee  
All other components vanish. The corresponding Riemann tensor has  
non-zero components  
               \be  
               \ba{l}  
               \dsp{ R_{uxux} = -\frac{1}{2}\, K_{,xx}, \hspace{2em}  
                     R_{uyuy} = -\frac{1}{2}\, K_{,yy}, }\\ 
                \\ 
               \dsp{ R_{uxuy} = R_{uyux} = - \frac{1}{2}\, K_{,xy}.}  
               \ea 
               \label{4} 
               \ee  
The only non-vanishing component of the Ricci tensor then is  
               \be  
               R_{uu} = - \frac{1}{2}\, \lh K_{,xx} + K_{,yy} \rh \equiv  
                 - \frac{1}{2}\, \Del_{trans} K. 
               \label{5} 
               \ee 
Here the label $trans$ refers to the transverse $(x,y)$-plane, with the  
$z$-axis representing the longitudinal direction. In complex notation  
               \be  
               \zg = x + iy, \hspace{3em} \bar{\zg} =  x - iy, 
               \label{6} 
               \ee 
the Einstein equations in vacuo become  
               \be  
               R_{\mu\nu} = 0 \hspace{2em} \Leftrightarrow \hspace{2em}  
                 K_{,\zg \bar{\zg}} = 0.  
               \label{7} 
               \ee 
The general solution of this equation reads  
               \be 
               K(u,\zg,\bar{\zg}) = f(u;\zg) + \bar{f}(u;\bzg) =
                \sum_{n=0}^{\infty}\, \int_{-\infty}^{\infty}  
                \frac{dk}{2\pi}\, \lh \eps_n(k) e^{-iku} \zg^n +  
                \bar{\eps}_n(k) e^{iku} \bar{\zg}^n \rh.  
               \label{8} 
               \ee 
Note that the terms with $n = 0,1$ correspond to vanishing Riemann tensor:  
$R_{\mu\nu\kg\lb} = 0$; therefore they represent flat Minkowski space-time in  
a non-standard choice of co-ordinates. For this reason we adopt the convention  
that $\eps_0 = \eps_1 = 0$, which is just a choice of gauge.  
\vs{3} 
                
\nit 
{\bf b.\ Geodesics of planar-wave space-times}  
\vs{1}  
 
\nit   
We proceed to solve the geodesic equation in the gravity-wave space-time  
(\ref{1}) along the lines of ref.\ct{jw2}:  
               \be  
               \ddot{x}^{\mu}\, +\, \Gam_{\nu\lb}^{\;\;\;\;\mu} \dot{x}^{\nu}  
                  \dot{x}^{\lb}\, =\, 0 
               \label{9} 
               \ee 
Here the overdot denotes a proper-time derivative. The proper-time Hamiltonian  
satisfies a constraint imposed by eq.(\ref{1}): 
               \be  
               \ba{lll} 
               H & = & \dsp{ g_{\mu\nu} \dot{x}^{\mu} \dot{x}^{\nu} } \\ 
                 & & \\ 
                 & = & \dsp{ -\,\dot{u} \dot{v}\, -\,  
                 K(u,x,y)\, \dot{u}^2\, +\, \dot{x}^2\, +\, \dot{y}^2\, =\,  
                 - c^2. } 
               \ea 
               \label{10} 
               \ee  
Because the metric is covariantly constant, the hamiltonian is a constant of  
motion:  
               \be 
               \dot{H} = 0.  
               \label{11} 
               \ee 
This can be checked directly from the geodesic equation (\ref{9}). Also, as  
$v$ is a cyclic co-ordinate, its conjugate momentum is conserved: 
               \be 
               \ddot{u}\, =\, 0,  
               \label{12} 
               \ee  
with the simple solution $\dot{u} \equiv \gam =$ constant. Again, this agrees  
with the geodesic equation, as there is no non-vanishing connection component  
in the $u$-direction: $\Gam_{\nu\lb}^{\;\;\;\;u} = 0$.    
            
\nit     
Only the equations of motion in the $x$-$y$-plane depend on the specific wave  
potential $K(u,x,y)$:  
               \be  
               \ba{lll}  
               \ddot{x} & = & \dsp{ - \frac{1}{2}\, K_{,x}\, \dot{u}^2\, =\,  
                 - \frac{\gam^2}{2}\, K_{,x}, }\\ 
                & & \\ 
               \ddot{y} & = & \dsp{ - \frac{1}{2}\, K_{,y}\, \dot{u}^2\, =\,  
                 - \frac{\gam^2}{2}\, K_{,y}, }  
               \ea  
               \label{13} 
               \ee  
Eqs.\ (\ref{10})-(\ref{13}) specify completely the motion of a test particle,  
with the conservation of $H$ taking the place of the equation for the  
acceleration in the $v$-direction: 
               \be 
               \gam\, \dot{v}\, +\, \gam^2 K(u,x,y)\, =\, \dot{x}^2\, +\,  
                 \dot{y}^2\, +\, c^2.  
               \label{14} 
               \ee  
If we now add $\dot{z}^2$ to the left- and right-hand side, and remember that  
               \be 
               \gam \dot{v} = \dot{u} \dot{v} = c^2 \dot{t}^2 - \dot{z}^2,  
               \label{15} 
               \ee  
we can rewrite the hamiltonian conservation law as  
               \be  
               c^2 \dot{t}^2 + \gam^2 K = c^2 + {\bf \dot{r}}^2.  
               \label{16} 
               \ee  
Finally, with ${\bf v} = d{\bf r}/dt = {\bf \dot{r}}/\dot{t}$, the equation  
can be cast into the form  
            \be 
            \dot{t}\, =\, \frac{dt}{d\tau}\, =\, \sqrt{\frac{1 - \gam^2 K/c^2}{ 
                        1 - {\bf v}^2/c^2}}  
            \label{17} 
            \ee  
This equation describes relativistic time-dilation as resulting from two  
effects: \nl  
(i) the usual special-relativistic time-dilation from the relative motion of  
observers in the rest- and laboratory frame, whose time co-ordinates are  
$\tau$ and $t$, respectively; \nl  
(ii) the gravitational redshift resulting from the non-trivial potential $K$. \nl  
Now from the conservation of $\gam = \dot{u} = c \dot{t} - \dot{z}$ it follows,  
that 
               \be 
               \gam = c \dot{t}\, \lh 1 - \frac{v_z}{c} \rh,  
               \label{18}  
               \ee  
with $v_z = dz/dt$. Eqs.\ (\ref{17}), (\ref{18}) can then be solved for $\gam$:  
           \be  
           \frac{\gam^2}{c^2}\, =\, \frac{1}{K + \dsp{ \frac{1 - {\bf v}^2/c^2} 
                {\lh 1 - v_z/c \rh^2}}}. 
           \label{19} 
           \ee   
Thus, for a paticle starting at rest at infinity in an asymptotically  
minkoskian space-time, we find $\gam = c$. At the same time we observe that  
               \be  
               h = K + \frac{1 - {\bf v}^2/c^2}{\lh 1 - v_z/c \rh^2} 
               \label{20}  
               \ee  
is conserved. Now we recall that in our conventions $K$ is at least quadratic  
in the transverse co-ordinates; hence the components $\ddot{x}$ and 
$\ddot{y}$ of the transverse acceleration vanish for $x = y =0$. 
Furthermore $K(u,0,0) = 0$, with the result that the origin of the 
transverse plane moves at constant velocity along the $z$-axis:  
\be  
\frac{\gam^2}{c^2}\, =\, \frac{1 - v_z/c}{1 + v_z/c} 
\hspace{1em} \Leftrightarrow \hspace{1em}  
\frac{v_z}{c}\, =\, \frac{1 - \gam^2/c^2}{1 + \gam^2/c^2}.  
\label{21}  
\ee 
In particular, the point at rest in the origin moves along the simple geodesic  
               \be 
               x^{\mu}(\tau) = (c\tau, 0, 0, 0).  
               \label{22} 
               \ee  
Taking this geodesic as our reference, the solution for the geodesic motion  
$\bar{x}^{\mu}(\tau)$ of any other test particle at the same time presents  
a measure for the geodesic deviation between the worldlines of the two particles.   
                
\section{Einstein-scalar waves} 
                
Having discussed the planar gravitational waves (\ref{1}) in empty space  
we now turn to discuss similar unidirectional wave solutions of the combined  
system of Einstein gravity and a set of massless self-interacting scalar  
fields. The solutions of the inhomogeneous and non-linear Einstein  
equations, with the energy-momentum tensor that of the right- (or left-)  
moving scalar waves, nevertheless turn out to be a linear superposition of  
the gravitational field of the scalar waves and the free gravitational  
wave solutions discussed in the first paragraph.  
    
We introduce a set of massless scalar fields $\sg^{i}(x)$, $i = 1,...,N$,  
taking values in a manifold with the dimensionless metric $G_{ij}[\sg]$.  
In four-dimensional space-time the fields themselves have dimension $[\sg]  
= \sqrt{E/l}$; thus, introducing an appropriate length scale $1/f$, in the 
context of quantum field theory we could write $\sg^{i} = \sqrt{\hbar c}\,  
f\, \eta^{i}$, with $\eta^{i}(x)$ a dimensionless field.  
 
The starting point of our analysis is given by the gravitational and 
$\sg$-model field equations  
          \be  
          \ba{l} 
          \Box^{cov}\, \sg^{i}\, +\, \Gam_{jk}^{\;\;\;\;i}[\sg]\, 
          g^{\mu\nu} \pl_{\mu} \sg^j \pl_{\nu} \sg^k\, =\,  0, \\ 
          \\  
          R_{\mu\nu}\, =\, \dsp{ -\, \frac{8\pi G}{c^4}\, G_{ij}[\sg]\,  
          \pl_{\mu}\sg^i \pl_{\nu} \sg^j. } 
          \ea  
          \label{2.1} 
          \ee  
Here the covariant d'Alembertian is defined on scalar fields in the standard  
fashion 
\[ \Box^{cov}\, =\, \frac{1}{\sqrt{-g}}\, \pl_{\mu}\, \sqrt{-g} g^{\mu\nu}\,  
   \pl_{\nu}, \] 
whilst $\Gam_{ij}^{\;\;\;\;k}[\sg]$ denotes the Riemann-Christoffel  
connection in the target manifold of the scalar fields. These equations can 
be derived straightforwardly from the combined Einstein-$\sg$-model action, 
but we will skip the details of that procedure here. Our aim is to construct 
simultaneous traveling wave solutions of the full set of equations 
(\ref{2.1}). Such solutions are actually quite easy to find. First,  
the scalar field equation is solved by taking right-moving fields  
\be  
\sg^{i} = \sg^{i}(u),  
\label{2.2}  
\ee  
with no dependence on any other co-ordinate. Next we substitute this  
solution of the scalar field into the second equation for the  
corresponding gravitational field. As before, only the $uu$-component  
of this equation survives, reading  
\be  
    R_{uu}\, =\, -\, \frac{1}{2}\, \Del_{trans} K\, =\, -\, \frac{8\pi G}{c^4}\,  
    G_{ij}[\sg]\, \pl_u \sg^i \pl_u \sg^j. 
    \label{2.3} 
    \ee  
As this is a linear equation, the general solution consists of a linear  
superposition of a particular solution and the general free gravitational 
wave of the previous section:  
\be 
   K(u,\zg,\bar{\zg})\, =\, \frac{8\pi G}{c^4}\, G_{ij}[\sg]\,  
   \pl_u \sg^i \pl_u \sg^j \bar{\zg} \zg\, +\, f(u,\zg)\, +\,  
   \bar{f}(u,\bar{\zg}).  
   \label{2.4}  
   \ee  
Now any specific solution $\sg^{i}(u)$ is a map from the real line into the  
target manifold of the scalar fields. Consider the special case that this
curve in the target manifold is a geodesic:
\be
\frac{d^2\sg^{i}}{du^2}\, +\, \Gam_{jk}^{\;\;i}\, \frac{d\sg^{j}}{du}\,
  \frac{d\sg^{k}}{du}\, =\, 0. 
\label{2.4.0}
\ee 
Then the quantity  
\be  
I\, =\, G_{ij}[\sg]\, \frac{d\sg^{i}}{du} \frac{d\sg^{j}}{du},  
\label{2.4.1} 
\ee  
generating translations in $u$, is constant along this curve: $dI/du = 0$.  
Moreover, for Euclidean manifolds with non-degenerate metric it is positive  
definite: $I > 0$. Observe, that for manifolds with compact directions 
(like spheres) the geodesics may be closed; then the corresponding scalar  
field configurations are periodi. 
 
The special solution for the accompanying gravitational field now  
becomes  
\be  
K_{scalar}(u,x,y)\, =\, \frac{4\pi G I}{c^4}\, (x^2 + y^2), 
\label{2.7} 
\ee  
to which an arbitrary free gravitational wave solution can be added.  
In this special case, upon inserting $K_{scalar}$ into eqs.(\ref{13})  
the transversal equations of motion of a test mass take the particularly  
simple form:  
\be  
\ddot{x}\, =\, -\, \frac{4\pi G I\gam^2}{c^4}\, x, \hspace{3em}  
\ddot{y}\, =\, -\, \frac{4\pi G I\gam^2}{c^4}\, y. 
\label{2.8} 
\ee  
Thus the test mass executes a simple harmonic motion in the transverse  
plane, with frequency  
\be  
\og\, =\, \frac{\gam}{c^2}\, \sqrt{4\pi GI}.  
\label{2.9} 
\ee  
The solutions for the coupled Einstein-scalar field equations discussed  
here are not the only ones of interest. For example, the gravitational  
waves accompanying expanding domain walls in a theory with a spontaneously  
broken global symmetry can be calculated and have been discussed e.g.\ in  
\ct{jw1,jw2}. 
 
\section{Einstein-Dirac waves}  
 
In this section we construct wave-solutions for massless chiral fermions  
coupled to Einstein gravity. As before the waves are unidirectional,  
and both left- and righthanded fermion solutions, associated with  
helicity $\pm 1$ quantum states, exist.  
 
To treat fermions in interaction with gravity, it is necessary to introduce  
the vierbein and spin connection into the formalism. With the local  
minkowski metric $\eta =$ diag$(+1,+1,+1,-1)$, the vierbein is a local 
lorentz vector of 1-forms $E^{a}(x) = dx^{\mu} e_{\mu}^{\;a}(x)$ satisfying 
the symmetric product rule  
\be  
\eta_{ab}\, E^{a} E^{b}\,  
=\, \eta_{ab}\, e_{\mu}^{\;a} e_{\nu}^{\;b} 
   dx^{\mu} dx^{\nu}\, =\, g_{\mu\nu} dx^{\mu} dx^{\nu}.   
\label{4.1} 
\ee  
In a convenient local lorentz gauge, the vierbein corresponding to the  
metric (\ref{1}) takes the form 
\be  
E^{a}\, =\, \lh dx, dy, \frac{1}{2}\, ((K - 1)\, du + dv),  
 \frac{1}{2}\, ((K + 1)\, du + dv) \rh.  
\label{4.2}  
\ee  
The inverse vierbein is defined by the differential operator  
$\nabla_a = e_a^{\, \mu} \pl_{\mu}$ such that  
\be 
E^a \nabla_a\, =\, dx^{\mu} \pl_{\mu}  
\label{4.3} 
\ee  
In components it reads  
\be  
\nabla_a\, =\, \lh \pl_x, \pl_y, -\pl_u + (K+1)\,\pl_v,  
 \pl_u - (K-1)\, \pl_v \rh.  
\label{4.4} 
\ee  
Next we compute the components of the spin connection $\og^a_{\; b}  
= dx^{\mu} \og_{\mu\;\;b}^{\;\;a}$ from the identity  
\be  
dE^a\, =\, \og^a_{\; b}\, \wedge\, E^b.  
\label{4.5} 
\ee  
With the vierbein (\ref{4.2}) the spin connection has only one component 
\be  
\og_u^{\; ab}\, =\, - \og_u^{\; ba}\, =\, \frac{1}{2}\,  
 \lh \ba{cccc} 0 & 0 & K_{,x} & K_{,x} \\ 
               0 & 0 & K_{,y} & K_{,y} \\ 
              -K_{,x} & -K_{,y} & 0 & 0 \\  
              -K_{,x} & -K_{,y} & 0 & 0 \\ \ea \rh.  
\label{4.6} 
\ee  
In order to construct the dirac operator we introduce a basis for  
the dirac matrices satisfying $\left\{ \gam^a, \gam^b \right\} = 2  
\eta^{ab}$, and define a set spinor generators for the lorentz algebra by  
$\sg_{ab} = \frac{1}{4}\, [\gam_a, \gam_b]$. Then the dirac  
operator is  
\be 
\gam \cdot D\, =\, \gam^a\, \lh \nabla_a - \frac{1}{2}\, \og_a^{\; bc} 
  \sg_{bc} \rh,  
\label{4.7} 
\ee  
The results we need all depend on the property of the light-cone components  
of the dirac algebra: 
\be 
\gam^u\, =\, \gam^a e_a^{\;u}\, =\, - \gam_3 + \gam_0. 
\label{4.7.1} 
\ee  
This element of the dirac algebra is nilpotent: 
\be  
(\gam^u)^2\, =\, 0. 
\label{4.7.2} 
\ee  
The same is true for $\gam_v = e_v^{\;a} \gam_a = \frac{1}{2} \gam^u$.  
Because of the form of the spin connection (\ref{4.6}), the dirac-algebra  
valued form $\og^{ab} \sg_{ab}$ is itself proportional to $\gam^u$; its 
nilpotency then guarantees that the spin-connection term in the covariant  
derivative (\ref{4.7}) vanishes by itself:  
\be  
\gam^a \og_a^{\;bc}\, \sg_{bc}\, =\, \gam^u  
   \og_u^{\; bc}\, \sg_{bc}\, =\, 0.  
\label{4.8} 
\ee  
Hence the only vestige of curved space-time left in the dirac operator  
is the inverse vierbein in the contraction of dirac matrices and  
differential operators:  
\be  
\ba{l}  
\gam \cdot D\, =\, \gam^a \nabla_a\, =\, \gam^{\mu} \pl_{\mu} \\ 
 \\ 
~~~ =\, \dsp{ i\, \lh \ba{cc}  
     \pl_u - (K - 1)\, \pl_v & -\sg_1 \pl_x - \sg_2 \pl_y \\ 
                             & - \sg_3\, (-\pl_u + (K+1)\, \pl_v) \\  
    \\ 
    \sg_1 \pl_x + \sg_2 \pl_y & -\pl_u + (K -1)\, \pl_v \\ 
    + \sg_3\, (-\pl_u + (K+1)\, \pl_v) & 
    \ea \rh. } 
\ea 
\label{4.9} 
\ee  
Here we have introduced the following basis for the dirac algebra: 
\be  
\gam_k\, =\, \lh \ba{cc} 0 & -i \sg_k \\ 
                        i \sg_k & 0 \ea \rh,\, k=1,2,3; 
 \hspace{2em} \gam^0\, =\, \lh \ba{cc} i 1 & 0 \\  
                                        0   & -i 1 \ea \rh,  
\label{4.9.0} 
\ee 
with the $\sg_k$ the standard pauli matrices. The zero modes of  
this operator with the property that the energy-momentum tensor only  
has a non-zero $T_{uu}$ component are flat spinor fields $\psi(u)$ of
the form
\be  
\psi(u)\, =\, i\, \gam^{u}\, \lh \ba{c} \chi(u) \\  
                                        0 \ea \rh\, =\, 
              \lh \ba{cc} 1 & -\sg_3 \\  
                         -\sg_3 & 1 \ea \rh\, \lh \ba{c} \chi(u) \\  
                                                           0 \ea \rh\,  
              =\, \lh \ba{c} \chi(u) \\ -\sg_3 \chi(u) \ea \rh,   
\label{4.9.1} 
\ee 
where $\chi(u)$ is a 2-component (pauli) spinor. Indeed, first of all  
spinors of this type are zero-modes of the dirac operator: 
\be  
\gam \cdot D \psi = 0. 
\label{4.9.2} 
\ee  
This follows by direct application of the expression (\ref{4.9}) to the  
spinor (\ref{4.9.1}), using the nilpotency of $\gam^u$. Moreover, with  
this property it also follows that the energy-momentum tensor takes the form  
\be  
T_{\mu\nu}\, =\, \frac{1}{8}\, \overline{\psi}\, \lh \gam_{\mu} D_{\nu}  
 + \gam_{\nu} D_{\mu} \rh\, \psi\, =\, \frac{1}{4}\, \del_{\mu}^u  
 \del_{\nu}^u\, \overline{\psi}\, \gam_u\, \pl_u\, \psi.  
\label{4.10} 
\ee  
To see this, first note that the $u$-component of the covariant derivative  
$D_{\mu}$ is the only one that does not vanish on $\psi(u)$ in general.  
We then only have to check that in all remaining cases with $\gam_{\mu}  
\neq \gam_u$ the spinor $\psi(u)$ (\ref{4.9.1}) gets multiplied by a  
dirac matrix which can be factorized such as to have a right multiplicator  
of the form $\gam^u$. Again, as $(\gam^u)^2 = 0$, $T_{\mu\nu}$ necessarily  
is of the required form (\ref{4.10}).  
 
Finally we remark, that the upper- and lower component of the pauli  
spinor $\chi(u)$ in our conventions correspond to a negative and positive  
helicity state, respectively. Thus we find as solutions of the dirac  
operator in the metric (\ref{1}) two massless spinor states, corresponding  
to right-moving zero-modes of the dirac operator with helicity $\pm 1$,  
respectively.  
 
This solution is self-consistent as the only non-zero component of the  
energy-momentum tensor is  
\be 
T_{uu}(u)\, =\, - \frac{1}{2}\, \left[ \chi^{\dagger} \chi^{\prime}  
  \right](u),  
\label{4.11} 
\ee 
where the prime denotes a derivative w.r.t.\ $u$, and the dagger on  
$\chi$ indicates hermitean conjugation of the 2-component spinor. It  
is then straightforward to solve the Einstein equation for $K$ in the  
presence of the energy momentum distribution of the spinor field: 
\be  
K_{spinor}(u,x,y)\, =\, -\, \frac{2\pi G}{c^4}\, \left[ \chi^{\dagger}  
 \chi^{\prime} \right](u)\, \lh x^2 + y^2 \rh.  
\label{4.12} 
\ee  
Again, to this particular solution an abitrary free gravitatonal wave  
can be added. It should be mentioned here, that consistency requires  
the spinors in the energy momentum tensor (\ref{4.10}), (\ref{4.11})  
to be anti-commuting objects, i.e.\ if the spinor fields $\chi(u)$  
are expanded in a fourier series of massless matter waves, the  
co-efficients take values in an infinite-dimensional Grassmann  
algebra. Thus the expression can be given an operational meaning  
only in the context of quantum theory, by performing some averaging  
procedure. For example, if the spinors form a condensate such that  
the kinetic energy $\Sg \equiv - \langle \left[\chi^{\dagger}  
\chi^{\prime} \right] \rangle$ = constant $> 0$, then such a condensate  
would generate gravitational waves in which test-masses perform  
harmonic motion of the type (\ref{2.8}), (\ref{2.9}) with frequency  
\be  
\og\, =\, \frac{\gam}{c^2}\, \sqrt{2 \pi G \Sg}.  
\label{4.13} 
\ee  
 
\section{Einstein-Maxwell waves}  
                
As the last example we consider coupled Einstein-Maxwell fields. We look  
for solutions of wave-type, using the metric (\ref{1}). In the absence  
of masses and charges, the field equations are:  
\be 
R_{\mu\nu}\, =\, -\, \frac{8\pi \ve_0 G}{c^2}\, \lh  
  F_{\mu\lb} F_{\nu}^{\;\;\lb} - \frac{1}{4} g_{\mu\nu} F^2 \rh,  
  \hspace{2em} D_{\lb} F^{\lb\mu} = 0. 
\label{3.1} 
\ee 
With the same metric (\ref{1}), we also find the same expressions for  
the components of the connection (\ref{3}), and the Riemann and Ricci  
curvature tensors (\ref{4}), (\ref{5}). Therefore the left-hand side  
of the Einstein eqn.\ (\ref{3.1}) is fixed in terms of the potential  
$K(u,y,z)$.  
 
As concerns the Maxwell equations, the covariant derivative  
\be 
D_{\lb} F^{\lb\mu} = \pl_{\lb} F^{\lb\mu} + \Gam_{\lb\nu}^{\;\;\;\;\lb} 
  F^{\nu\mu} + \Gam_{\lb\nu}^{\;\;\;\;\mu} F^{\lb\nu}  
\label{3.2} 
\ee  
reduces to the first term on the r.h.s., an ordinary four-divergence;  
this happens because in the last term the even connection is contracted  
with the odd field-strength tensor, whilst the middle term contains  
a trace over an upper and a lower index of the connection, which  
vanishes in our case.  
 
Thus the Maxwell equation reduces to the same expression as in  
minkowski space-time, and it has the same wave solutions. We consider  
an elementary wave solution, which in terms of the co-ordinate system  
(\ref{1}) is described by the vector potential  
\be 
A_{\mu} = ({\bf a} \sin k(ct - z), 0, 0), 
\label{3.3} 
\ee 
with the light-cone components vanishing, and with ${\bf a}$ a 
constant transverse vector: $a_z = 0$. Of course, arbitrary solution 
can be constructed from the elementary waves (\ref{3.3}) by linear  
superposition. With $u = ct - z$ and $\og = kc$ the angular  
frequency of the wave, the electric and magnetic fields are  
\be  
{\bf E}_k(u) = \og {\bf a} \cos ku, \hspace{2em}  
  {\bf B}_k(u) = {\bf k} \times {\bf a}\, \cos ku.  
\label{3.4} 
\ee  
As usual for e.m.\ waves, $|{\bf E}_k(0)| = c|{\bf B}_k(0)|$,  
and ${\bf E}_k \cdot {\bf B}_k = 0$. Indeed, the only non-zero  
components of the full field strength are  
\be 
F_{ui} = - F_{iu} = k a_i \cos ku, \hspace{3em} i = (x,y), 
\label{3.5} 
\ee  
all others vanishing. It is now straightforward to compute the  
stress-energy tensor components of the electro-magnetic field,  
with the result  
\be 
T_{uu} = \ve_{0} c^2 k^2 {\bf a}^2 \cos^2 ku,  
\label{3.6} 
\ee  
and all other components zero. The Einstein-Maxwell equations  
then reduce to 
\be  
\Del_{trans} K = \frac{16 \pi \ve_0 G}{c^2}\,  
   k^2 {\bf a}^2 \cos^2 ku. 
\label{3.7} 
\ee 
This has the special solution 
\be  
K_{em} = \frac{4 \pi \ve_0 G}{c^2}\,  
 k^2 {\bf a}^2 \cos^2 ku\, \lh x^2 + y^2 \rh\,=  
 \frac{4 \pi \ve_0 G}{c^4}\, {\bf E}_k^2(u)\, \bzg \zg. 
\label{3.8} 
\ee  
In view of the linearity of eq.(\ref{3.7}), the general solution  
is a superposition of such special solutions and arbitrary free  
gravitational waves of the type (\ref{8}): 
\be 
K(u,\zg,\bzg) = K_{em}(u,\zg,\bzg) + f(u,\zg) + \bar{f}(u,\bzg). 
\label{3.9} 
\ee  
\vs{1}  
 
\nit 
Next we turn to the motion of a test particle with mass  
$m$ and charge $q$ in the background of these gravitational and  
electro-magnetic fields. These equations are modified to take into  
account the Lorentz force on the test charge: 
\be 
\ddot{x}^{\mu} + \Gam_{\nu\lb}^{\;\;\;\;\mu} \dot{x}^{\nu}  
  \dot{x}^{\lb} = \frac{q}{m}\, F^{\mu}_{\;\;\nu} \dot{x}^{\nu}.  
\label{5.1} 
\ee  
With the only non-zero covariant components of $F_{\mu\nu}$ given  
by eq.(\ref{3.5}), there are no contravariant components in the  
lightcone direction $u$. As a result the equation for $u$ is not  
modified, and we again find 
\be 
\dot{u} = \gam = \mbox{const.} 
\label{5.2} 
\ee  
This also follows, because the electro-magnetic forces do not  
change the proper-time hamiltonian:  
\be  
\ba{lll} 
H & = & \dsp{ g_{\mu\nu} \dot{x}^{\mu} \dot{x}^{\nu} } \\ 
  & & \\ 
  & = & \dsp{ =\, -\,\dot{u} \dot{v}\, -\,  
  K(u,y,z)\, \dot{u}^2\, +\, \dot{y}^2\, +\, \dot{z}^2\, =\,  
  - c^2, } 
\ea 
\label{5.3} 
\ee  
except that $K(u,x,y)$ now is given by the modified expression  
(\ref{3.9}). Therefore $v$ is still a cyclic co-ordinate and  
equation (\ref{14}) for $\dot{v}$ again follows from the  
conservation of $H$:  
\be 
\gam\, \dot{v}\, +\, \gam^2 K(u,x,y)\, =\, \dot{x}^2\, +\,  
  \dot{y}^2\, +\, c^2.  
\label{5.4} 
\ee  
As a result we find in this case the same formal expressions  
for the solution of the equations of motion in the time-like  
and longitudinal directions:  
\be 
\dot{t}\, =\, \frac{dt}{d\tau}\, =\, \sqrt{\frac{1 - \gam^2 K/c^2}{ 
         1 - {\bf v}^2/c^2}},  
\label{5.5} 
\ee  
whilst  
\be  
h = K + \frac{1 - {\bf v}^2/c^2}{\lh 1 - v_z/c \rh^2} 
\label{5.6}  
\ee  
is again a constant of motion. In both cases of course $K$ now is the  
full solution (\ref{3.9}). 
         
Manifest changes in the equations of motion are obtained in the  
transverse directions:  
\be  
\ddot{\bfxi}\, =\, - \frac{\gam^2}{2}\, \mbox{{\boldmath $\nabla$}}_{\xi} K\, -\,  
  \frac{q \gam}{m}\, k {\bf a} \cos ku, 
\label{5.7} 
\ee  
where $\bfxi = (x,y)$ is a transverse vector and {\boldmath $\nabla$}$_{\xi}$  
is the gradient in the transverse plane. If we take for $K$ the  
special solution (\ref{3.8}), we find the conservation law  
\be  
\frac{4 \pi \ve_0 G}{c^2}\, k^2 {\bf a}^2 \bfxi^2\, \cos^2 ku\,  
 +\, \frac{1 - {\bf v}^2/c^2}{\lh 1 - v_z/c \rh^2}\, =\,  
 h\, =\, \mbox{const.} 
\label{5.8} 
\ee 
Inserting the explicit form of $u(\tau) = \gam \tau$,  
eqs.(\ref{5.7}) then take the form  
\be  
\ddot{\bfxi}\, =\, -\, \frac{4\pi\ve_0 G}{c^2}\, \gam^2 k^2  
 {\bf a}^2 \cos^2 (\gam k \tau)\, \bfxi\, -\, \frac{q \gam}{m}\,  
 k {\bf a} \cos (\gam k \tau). 
\label{5.9} 
\ee  
Equivalently, we can use $u$ instead of $\tau$ as the independent 
variable: 
\be  
\frac{d^2\bfxi}{du^2}\, =\, -\, \frac{4\pi\ve_0 G}{c^2}\, k^2  
 {\bf a}^2 \cos^2 (ku)\, \bfxi\, -\, \frac{q}{m\gam}\,  
 k {\bf a} \cos ku. 
\label{5.10} 
\ee   
Clearly, it is useful to decompose $\bfxi$ into components parallel  
and orthogonal to the electric field ${\bf E}_k$, which in our  
choice of electro-magnetic gauge is the same as that of the vector  
potential ${\bf a}$:  
\be  
\bfxi = \bfxi_{\parallel} + \bfxi_{\perp},  
\label{5.11} 
\ee 
with  
\be  
\bfxi_{\parallel} = \frac{\bfxi \cdot {\bf a}}{|{\bf a}|^2}\, {\bf a},  
 \hspace{3em}  
\bfxi_{\perp} = \frac{\bfxi \times {\bf a}}{|{\bf a}|}. 
\label{5.12} 
\ee  
It follows that  
\be  
\ba{lll} 
\dsp{ \frac{d^2\bfxi_{\parallel}}{du^2} } & = & \dsp{ -\,  
 \frac{4\pi\ve_0 G}{c^2}\, k^2 {\bf a}^2 \cos^2 (ku)\,  
 \bfxi_{\parallel}\, -\, \frac{q}{m\gam}\, k {\bf a} \cos ku, }\\ 
  & & \\ 
\dsp{ \frac{d^2\bfxi_{\perp}}{du^2} } & = & \dsp{ -\,  
 \frac{4\pi\ve_0 G}{c^2}\, k^2 {\bf a}^2 \cos^2 (ku)\, \bfxi_{\perp}. } 
\ea  
\label{5.13} 
\ee   
Transforming to the cosine of the double argument, the last equation 
can be seen to reduce to the standard Mathieu equation:  
\be  
\frac{d^2\bfxi_{\perp}}{du^2}\, +\, \frac{2\pi\ve_0 G}{c^2}\,  
  k^2 {\bf a}^2 \lh 1 + \cos 2ku \rh\, \bfxi_{\perp}\, =\, 0,  
\label{5.14} 
\ee 
whilst the other equation becomes an inhomogeneous Mathieu 
equation, with the Lorentz force representing the inhomogeneous 
term:  
\be  
\frac{d^2\bfxi_{\parallel}}{du^2}\, +\, \frac{2\pi\ve_0 G}{c^2}\,  
  k^2 {\bf a}^2 \lh 1 + \cos 2ku \rh\, \bfxi_{\parallel}\, =\,  
  -\, \frac{q}{m\gam}\, k {\bf a} \cos ku.  
\label{5.15} 
\ee 
Obviously, one may try to find a particular solution to this  
equation by making an expansion in powers of $\cos ku$. The general  
solution is a superposition of this special one plus the  
general solution of the Mathieu equation (\ref{5.14}).  
 
A special case is that of {\em static} crossed electric and  
magnetic fields, obtained in the limit $k \rightarrow 0$.  
Then the eqs.(\ref{5.14}) and (\ref{5.15}) reduce to ordinary  
homogeneous and inhomogeneous harmonic equations:  
\be  
\ba{lll}  
\dsp{ \frac{d^2\bfxi_{\perp}}{du^2}\, +\, \frac{4\pi\ve_0 G}{c^4}\,  
  {\bf E}_0^2\, \bfxi_{\perp} } & = & 0, \\ 
  & & \\ 
\dsp{ \frac{d^2\bfxi_{\parallel}}{du^2}\, +\, \frac{4\pi\ve_0 G}{c^4}\,  
  {\bf E}_0^2\, \bfxi_{\parallel} } & = & \dsp{  
  -\, \frac{q}{mc\gam}\, {\bf E}_0. } 
\ea 
\label{5.16} 
\ee 
The angular frequency of this harmonic motion is  
\be 
\og\, =\, \sqrt{ \frac{4\pi \ve_0 G}{c^2} }\, E_0\, =\,  
  0.29 \times 10^{-18} E_0\, \mbox{(V/m)}. 
\label{5.17} 
\ee  
Clearly, the Lorentz force due to the constant electric field  
produces a constant proper-time acceleration of the test charge, but  
the harmonic gravitational component of the motion is very slow for  
practically realistic electric fields: periods of a year or less  
require a field strength of the order of $10^{10}$ V/m or more.  
\vs{2}

\end{document}